\documentclass[english,aps,manuscript]{revtex4}
\usepackage[T1]{fontenc}
\usepackage[latin1]{inputenc}
\usepackage{babel}
\usepackage{graphics}

\makeatletter

\providecommand{\LyX}{L\kern-.1667em\lower.25em\hbox{Y}\kern-.125emX\@}

\newcommand{\bee}{\begin{equation}}
\newcommand{\ee}{\end{equation}}
\newcommand{\beea}{\begin{eqnarray}}
\newcommand{\eea}{\end{eqnarray}}

\makeatother
\begin{document}
\noindent \hspace*{10cm} COLO-HEP-468\\
 \hspace*{10.5cm} June 2001.

\noindent \vspace*{1cm}

{\centering \textbf{\Large Two Flavors of Staggered Fermions with
Smeared Links }\Large \par}

{\centering \vspace{0.5cm}\par}

{\centering {\large Anna Hasenfratz\( ^{\dagger } \) and Francesco
Knechtli\( ^{\ddagger } \) }\large \par}

{\centering \vspace{0.5cm}\par}

{\centering Physics Department, University of Colorado, \\
 Boulder, CO 80309 USA\par}

{\centering \vspace{0.5cm}\par}

{\centering \textbf{Abstract}\par}

{\centering \vspace{0.5cm}\par}

{\centering Staggered fermions with smeared links can have greatly
improved chiral properties. In a recent paper we introduced a simple
and effective method to simulate four flavors of staggered smeared
link fermions. In this work we extend the four flavor method to two
flavors. We define the two flavor action by the square root of the
four flavor fermionic determinant and show that by using a polynomial
approximation the two flavor action can be evaluated with the necessary
accuracy. We test this method by studying the finite temperature phase
structure with hypercubic smeared (HYP) link staggered action on \( N_{t}=4 \)
temporal lattices. \par}

\vspace{0.5cm}


\vfill

\( ^{\dagger } \) {\small e-mail: anna@eotvos.colorado.edu}{\small \par}

\( ^{\ddagger } \) {\small e-mail: knechtli@pizero.colorado.edu}{\small \par}

\eject

\section{Introduction }

In a series of recent papers we introduced the hypercubic blocked
(HYP) action where the gauge links that connect the fermions are smeared
with an optimized local block transformation, the hypercubic blocking
\cite{Hasenfratz:2001hp}. We showed that even one level of hypercubic
blocking improves the flavor symmetry of staggered fermions by an
order of magnitude but it distorts the local properties of the configurations
only minimally. The static potential calculated with HYP blocked links,
for example, is indistinguishable from the thin link one at distances
\( r/a\geq 2 \). We proposed a simple but effective method to simulate
four-flavor dynamical staggered actions with smeared links and presented
the first results on the finite temperature phase structure with HYP
action in Ref. \cite{Hasenfratz:2001ef}.

Our thermodynamic results with the HYP action are very different from
the well known results obtained with thin link four flavor staggered
fermions but consistent with the predictions of the instanton liquid
model \cite{Schafer:1996pz}. The thin link action shows a pronounced
first order phase transition on \( N_{t}=4 \) temporal lattices.
Two state signals have been observed also on \( N_{t}=6 \) and 8
lattices though the discontinuity weakens as the temporal lattice
size increases. The critical temperature is estimated to be \( T_{c}\simeq 150-170\,  \)MeV
\cite{Engels:1997ag}. In contrast to that with the the HYP action
we did not find any sign of a first order phase transition neither
on \( N_{t}=4 \) nor on \( N_{t}=6 \) temporal lattices even with
very light quarks. The quark mass dependence of the chiral condensate
suggests that if a phase transition occurs at all in the chiral massless
limit it is at a much lower temperature value than the above quoted
\( T_{c} \) and simulations would require larger values of \( N_{t} \)
even with the HYP action. We argued that the difference is due to
the improved flavor symmetry of the HYP action and suggested that
the thin link phase transition with \( N_{t}=4-8 \) is not much more
than a lattice artifact. 

The simulation method for smeared actions presented and used in Ref.
\cite{Hasenfratz:2001ef} works with any fermionic formulation but
requires the (stochastic) evaluation of the fermionic action. That
can be done straightforwardly for four flavors of staggered or two
flavors of Wilson/clover fermions but requires further considerations
for one or two flavors of staggered fermions. In this work we approximate
the two flavor staggered fermion determinant with the square root
of the four flavor fermionic determinant and evaluate the square root
by using a polynomial approximation. Because the fermions couple to
the smooth HYP fat links even a low order (32-64) polynomial is sufficient
to evaluate the fermionic action accurately. We use this simulation
technique in a preliminary study of the finite temperature phase structure
of the two flavor staggered HYP fermions on lattices with \( N_{t}=4 \)
temporal extension.

Before describing our method to simulate the two-flavor HYP action
we would like to mention an issue general to any two and 2+1 flavor
simulations based on staggered fermions. The partition function of
a dynamical system with \( n_{f} \) number of fermion flavors is
\begin{equation}
\label{Z_fermion}
Z_{n_{f}}=\int DUD\bar{\psi }D\psi \, e^{-S_{G}(U)-\sum ^{n_{f}}_{i=1}\bar{\psi }^{i}\Omega (V,m)\psi ^{i}}
\end{equation}
 \begin{equation}
\label{Z_integrated}
=\int DU\, det(\Omega ^{n_{f}}(V,m))\, e^{-S_{G}(U)}
\end{equation}
where \( S_{G}(U) \) is the gauge action depending on the thin gauge
links \( U_{i,\mu } \) and \( \Omega (V,m) \) is the fermion matrix
for one fermionic flavor depending on the smeared links \( V \) and
quark mass \( m \). The dependence of the smeared links on the thin
links is arbitrary but deterministic. For \( n_{f}=4 \) flavors of
staggered fermions the fermionic matrix is\[
\Omega ^{4}(V,m)=M^{\dagger }M\]
 where \begin{equation}
\label{M_staggered}
M(V,m)_{i,j}=2m\delta _{i,j}+\sum _{\mu }\eta _{i,\mu }(V_{i,\mu }\delta _{i,j-\widehat{\mu }}-V^{\dagger }_{i-\widehat{\mu },\mu }\delta _{i,j+\widehat{\mu }})\, .
\end{equation}
Here \( \eta _{i,\mu } \) are the usual staggered fermion phases
and \( M^{\dagger }M \) is defined on even or odd sites only. The
naive generalization for arbitrary flavors suggests the fermionic
matrix \begin{equation}
\label{Om_nf}
\Omega ^{n_{f}}(V,m)=(M^{\dagger }M)^{n_{f}/4}
\end{equation}
and the partition function \begin{equation}
\label{Z_staggered_nf}
Z_{n_{f}}=\int DU\, det((M^{\dagger }M)^{n_{f}/4})\, e^{-S_{G}(U)}
\end{equation}
\begin{equation}
\label{Z_staggered_fermion}
=\int DUD\bar{\psi }D\psi e^{-S_{G}(U)-\sum ^{n_{f}}_{i=1}\bar{\psi }^{i}(M^{\dagger }M)^{1/4}\psi ^{i}}\, .
\end{equation}
While eq. (\ref{Z_staggered_nf}) is formally well defined, its physical
meaning is less obvious for \( n_{f} \) that is not the multiple
of 4. Even if \( M^{\dagger }M \) described four degenerate flavors,
its fractional power is a non-local quantity and the action in eq.
(\ref{Z_staggered_fermion}) is non-local. To make the matter even
worse, the staggered fermion action does not describe four degenerate
flavors. It has only a remnant U(1) flavor symmetry and only one Goldstone
pion. The symmetry structure of the fractional power of the staggered
fermion matrix is not at all clear. Nevertheless actions like the
one described in eqs. (\ref{Z_staggered_nf}-\ref{Z_staggered_fermion})
have been used extensively to study two and 2+1 flavor systems. One
might argue that in the continuum limit there must exist a local action
that describes a single fermion species and the fourth power of this
action should agree with the four flavor staggered action. In that
case the 1/4th power of the staggered fermion matrix can be considered
as a non-local approximation to the well defined one-flavor action.
Perturbation theory supports this argument \cite{Bernard:1994sv}.
At finite lattice spacing non-perturbative lattice effects can destroy
this approximate agreement and simulations with \( n_{f} \) staggered
flavors cannot be considered as true \( n_{f} \) flavor simulations.
How can one test if and to what accuracy eq. (\ref{Z_staggered_nf}-\ref{Z_staggered_fermion})
describe \( n_{f} \) degenerate dynamical species? There are two
sides of this question. First, even with \( n_{f}=4 \) when the lattice
fermion action is local, we have to ask if flavor symmetry is sufficiently
restored so the staggered action can be considered to describe four
degenerate flavors. Second we have to consider when it is a good approximation
to take a fractional power of the determinant to describe \( n_{f}=1 \)
or 2 flavors. One does not expect the \( n_{f}\neq 4 \) simulations
to be correct when the corresponding \( n_{f}=4 \) simulations are
not, but there is no guarantee that lattice artifacts do not increase
significantly when the fractional power of the determinant is taken. 

The properties of the vacuum give the cleanest signal to answer both
questions. For example the topological susceptibility at finite, small
quark mass is inversely proportional the the fermion number and the
proportionality constant is predicted by chiral perturbation theory.
Presently available data suggests that with thin link staggered fermions
the lattice has to be fairly smooth, \( a<0.1 \)fm to describe two
flavors of degenerate fermions. On lattices with \( a\approx 0.17 \)fm
both the two and four flavor simulations fail at reproducing the correct
chiral behavior. With the HYP action already at \( a\approx 0.17 \)fm
the topological susceptibility is close to its chiral value, at least
with four flavors \cite{Hasenfratz:2001wd}. 

We will not be concerned with the above outlined theoretical problem
concerning two flavor staggered fermions in this paper. We consider
the action eqs. (\ref{Z_staggered_nf}-\ref{Z_staggered_fermion})
and describe an effective and simple method to simulate it. Whether
this action describes the correct number of flavors has to be investigated
for different actions at different lattice spacings independently.

\section{Simulating two flavors of Staggered Fermions}

In Ref. \cite{Hasenfratz:2001ef} we simulated actions of the form
of eq. (\ref{Z_integrated}) using a two step algorithm. First a subset
of the thin links are updated using an over relaxation or heat bath
updating based on the pure gauge action \( S_{G}(U) \). The fermionic
determinant is not invariant under such an update and the proposed
change is accepted only with probability \begin{equation}
\label{pacc}
P_{acc}(V^{\prime },V)=\min \left\{ 1,\exp [-S_{F}(V^{\prime })+S_{F}(V)]\right\} \, ,
\end{equation}
 where \( V^{\prime } \) denotes the new fat link configuration and
\( S_{F}(V)=-tr\, ln(\Omega ^{n_{f}})) \) is the fermionic action.
If a given sequence of thin link updates is generated with the same
probability as the reverse sequence this procedure satisfies the detailed
balance condition. The fermionic action can be evaluated using a stochastic
estimator if the fermionic matrix has the form\begin{equation}
\label{Om}
\Omega ^{n_{f}}=Q^{\dagger }(V)Q(V)\, .
\end{equation}
 In that case the acceptance probability eq. (\ref{pacc}) can be
approximated stochastically as\begin{equation}
\label{stochacc}
P^{\prime }_{acc}(V^{\prime },V)=\min \left\{ 1,\exp \left\{ \xi ^{\dagger }\left[ Q^{\dagger }(V^{\prime })Q(V^{\prime })-Q^{\dagger }(V)Q(V)\right] \xi \right\} \right\} \, ,
\end{equation}
where the vector \( \xi  \) is generated according to the probability
distribution\begin{equation}
\label{xsivec}
P(\xi )\propto \exp \left\{ -\xi ^{\dagger }Q^{\dagger }(V^{\prime })Q(V^{\prime })\xi \right\} \, .
\end{equation}
 For \( n_{f}=4 \) flavors of staggered fermions the algorithm is
straightforward. With the substitution \( Q=M, \) where \( M \)
is the matrix defined in eq. (\ref{M_staggered}), the vector \( \xi  \)
can be written as \[
\xi =M^{-1}R\, ,\]
 where \( R \) is a Gaussian random vector and the action difference
is easily evaluated as \[
-S_{F}(V^{\prime })+S_{F}(V)=\xi ^{\dagger }\left[ M^{\dagger }(V^{\prime })M(V^{\prime })-M^{\dagger }(V)M(V)\right] \xi \, .\]

In order to simulate \( n_{f}=1 \) or 2 flavors we have to write
\( \Omega ^{n_{f}}=(M^{\dagger }M)^{n_{f/4}} \) in the form of eq.
(\ref{Om}). That can be done using a polynomial approximation for
fractional powers. In case of \( n_{f}=2 \) we need \begin{equation}
\label{poly_appr1}
x^{1/2}=P_{1/2}(x)=lim_{n\to \infty }P^{(n)}_{1/2}(x)
\end{equation}
 or \begin{equation}
\label{poly_appr}
x^{1/2}=x\, P_{-1/2}(x)=x\, lim_{n\to \infty }P^{(n)}_{-1/2}(x)\, ,
\end{equation}
 where \( P_{\pm 1/2}^{(n)}(x)=\sum ^{n}_{i=0}c_{i}^{(n)}x^{i} \)
is an \( n \)th order polynomial approximation of the function \( x^{\pm 1/2} \).
The form of eq. (\ref{poly_appr}) is more convenient to use as the
polynomial \( P_{-1/2}^{(n)}(x) \) can be chosen such that it has
only complex roots if \( n \) is even and can be written as \[
P_{-1/2}^{(n)}(x)=c^{(n)}_{n}\prod _{i=1}^{n/2}(x-r^{(n)}_{i})(x-r_{i}^{*(n)})=q^{(n)}_{-1/2}(x)q_{-1/2}^{\dagger (n)}(x)\, ,\]
 where \( r^{(n)}_{i} \) denotes the \( i \)th complex root with
positive imaginary part and \begin{equation}
\label{qn(x)}
q_{-1/2}^{(n)}(x)=\sqrt{c^{(n)}_{n}}\prod _{i=1}^{n/2}(x-r^{(n)}_{i})\, .
\end{equation}
 With this notation the fermion matrix for \( n_{f}=2 \) staggered
fermion flavors is \[
\Omega ^{2}=(M^{\dagger }M)^{1/2}=M^{\dagger }M\, P_{-1/2}(M^{\dagger }M)\]
\[
=lim_{n\to \infty }q_{-1/2}^{\dagger (n)}(M^{\dagger }M)\, M^{\dagger }M\, q^{(n)}_{-1/2}(M^{\dagger }M)\, ,\]
 where we have used the fact that \( q^{(n)}_{-1/2}(M^{\dagger }M) \)
commutes with \( M^{\dagger }M \). Identifying\begin{equation}
\label{Qmat}
Q=lim_{n\to \infty }M\, q^{(n)}_{-1/2}(M^{\dagger }M)
\end{equation}
 in eq. (\ref{Om}) allows us to simulate the two flavor staggered
action. The vector \( \xi  \) of eq. (\ref{xsivec}) can be generated
as \begin{equation}
\label{xi_nf2}
\xi =lim_{n\to \infty }q^{\dagger (n)}_{-1/2}(M^{\prime \dagger }M^{\prime })R\, ,
\end{equation}
 where \( R \) is a Gaussian random vector and the action difference
in eq. (\ref{stochacc}) can be calculated as\[
\Delta S=-S_{F}(V^{\prime })+S_{F}(V)\]

\begin{equation}
\label{delta_S}
=lim_{n\to \infty }\xi ^{\dagger }\left[ P^{(n)}_{-1/2}(M^{\prime \dagger }M^{\prime })\, M^{\prime \dagger }M^{\prime }-P^{(n)}_{-1/2}(M^{\dagger }M)\, M^{\dagger }M\right] \xi 
\end{equation}

\section{Implementation of the Polynomial Approximation}

We followed the method outlined in Ref. \cite{Montvay:1996ea} to
construct the polynomial approximation for the function \( x^{\alpha } \)
in the form \( P_{\alpha }^{(n)}(x)=\sum ^{n}_{i=0}c^{(n)}_{i}x^{i}. \)
To find the coefficients \( c^{(n)}_{i} \)we minimize the integral\[
I=\int ^{\lambda }_{0}dx\, (x^{\alpha }-P^{(n)}(x))^{2}x^{\omega }.\]
 The term \( x^{\omega } \) regularizes the \( x\approx 0 \) behavior
of the integral and has to be chosen such that \( 2\alpha +\omega \geq 0 \).
With the limits of the integral set to 0 and \( \lambda  \) the resulting
polynomial will approximate \( x^{\alpha } \) in the (0,\( \lambda  \))
interval. The minimization condition of the integral \( I \) leads
to \( n+1 \) linear equations that can be solved analytically. We
used Mathematica to determine the coefficients \( c_{i}^{(n)} \)
for \( \alpha =-1/2 \) at several \( \omega  \) values for polynomials
of order \( n=32-256 \). The roots of the polynomials can also be
determined using Mathematica. The roots \( r^{(n)}_{i} \) and the
coefficient \( c^{(n)}_{n} \) for several \( n \) values and \( \omega =1 \)
can be found on the web site http://www-hep.colorado.edu/\textasciitilde{}anna/Polynomials/.
The coefficients on the web site correspond to \( \lambda =1 \) and
have to be rescaled if a different \( \lambda  \) is required. 

In staggered fermion simulations we need to approximate \( (M^{\dagger }M)^{-1/2} \)
in the interval \( (0,\lambda ) \) that covers the eigenvalue spectrum
of the matrix \( M^{\dagger }M \). For free staggered fermions the
maximum eigenvalue is \( \lambda =16 \). In simulations we used \( \lambda =18 \)
to cover the occasional \( \lambda >16 \) eigenvalues that arise
due to fluctuations. 

In evaluating \( \xi  \) or \( \Delta S \) we need to multiply lattice
vectors by \( q^{(n)}_{-1/2} \) or \( q_{-1/2}^{\dagger (n)} \).
Each such operation involves \( n/2 \) multiplications by \( M^{\dagger }M \).
A typical example is \begin{equation}
\label{mult_1}
\Phi =q_{-1/2}^{(n)}(M^{\dagger }M)\varphi =\sqrt{c_{n}^{(n)}}\prod _{i=1}^{n/2}(M^{\dagger }M-r^{(n)}_{i})\varphi 
\end{equation}
 where \( \varphi  \) is an arbitrary source vector and \( \Phi  \)
is the resulting vector of the operation. To reduce numerical round-off
errors it is important to order the terms in the product such that
the norms of the intermediate lattice vectors do not fluctuate too
much \cite{Montvay:1997vh}. One such ordering is also given on the
above web site.  It is also helpful to factor out the scale \( \lambda  \)
and rewrite eq. (\ref{mult_1}) as\[
\Phi =\prod _{i^{\prime }=1}^{n/2}[(d^{(n)}\lambda )(\frac{M^{\dagger }M}{\lambda }-\frac{r_{i^{\prime }}^{(n)}}{\lambda })]\varphi \]
 where \( i^{\prime } \) denotes some chosen sequence of the original
index \( i \) and \( d^{(n)}=(c_{n}^{(n)})^{1/n} \). This way the
multiplication with \( q_{-1/2}^{(n)} \) can be performed without
excessive round-off errors even with \( n=256 \).

To simplify the notation in the following we will not write out the
argument \( M^{\dagger }M \) or \( M^{\prime \dagger }M^{\prime } \)
in the polynomials but use the notation \( q_{-1/2}\equiv q_{-1/2}(M^{\dagger }M) \),
\( q^{\prime }_{-1/2}\equiv q_{-1/2}(M^{\prime \dagger }M^{\prime }) \),
etc. 

By using a finite order polynomial approximation of \( q_{-1/2} \)
we introduce systematical errors. One can reduce these errors both
by increasing the order of the approximating polynomial and by improving
the operator used in evaluating \( \Delta S \) in eq. (\ref{delta_S}).
To see the latter let us denote the action difference evaluated according
to eq. (\ref{delta_S}) using an \( n \)th order polynomial by \( \overline{\Delta S}^{(n)} \)
and use the notation \( \Delta ^{(n)}=P_{-1/2}^{(n)}-P_{-1/2} \).
Now we can write the action difference as \begin{equation}
\label{delta_S_bar}
\overline{\Delta S}^{(n)}=\Delta S+\xi ^{\dagger }\left[ \Delta ^{\prime (n)}\, M^{\prime \dagger }M^{\prime }-\Delta ^{(n)}\, M^{\dagger }M\right] \xi \, .
\end{equation}
 Since \( \Delta S \) is small (otherwise the proposed change is
not accepted), \( M^{\prime \dagger }M^{\prime }-M^{\dagger }M \)
will also be small. That implies that the correction to \( \Delta S \)
is small if \( \Delta ^{(n)} \) is small. But we can do even better
by considering the quantity \[
\widetilde{\Delta S}^{(n)}=\xi ^{\dagger }\left[ \left( P_{-1/2}^{\prime (n)}\right) ^{3}\, \left( M^{\prime \dagger }M^{\prime }\right) ^{2}-\left( P^{(n)}_{-1/2}\right) ^{3}\, \left( M^{\dagger }M\right) ^{2}\right] \xi \]
\[
=\Delta S+3\xi ^{\dagger }\left[ \Delta ^{\prime (n)}\, M^{\prime \dagger }M^{\prime }-\Delta ^{(n)}\, M^{\dagger }M\right] \xi +O(\Delta ^{(n)2})\, .\]
The combination \[
\Delta S^{(n)}=\frac{1}{2}(3\overline{\Delta S}^{(n)}-\widetilde{\Delta S}^{(n)})=\Delta S+O(\Delta ^{(n)2})\]
has only \( O(\Delta ^{(n)2}) \) errors. A similar improvement is
possible for the vector \( \xi  \) but it is too cumbersome and we
have decided to use a higher order polynomial with the simple form
of eq. (\ref{xi_nf2}) instead.

\section{Summary of the \protect\( n_{f}=2\protect \) flavor Updating}

In this section we summarize the accept-reject step of our algorithm
that follows the sequence of over relaxation or heat bath updates
of the gauge links. The over relaxation and heat bath steps can be
performed the same way for two flavors as for four flavors and the
details are discussed in Ref. \cite{Hasenfratz:2001ef}. 

Once the thin links are updated and the smeared \( V^{\prime } \)
links are calculated the proposed change is accepted with a probability
given by eq. (\ref{pacc}). The acceptance probability is evaluated
using a stochastic estimator and in Ref. \cite{Knechtli:2000ku} we
showed that the acceptance rate can be increased significantly if
the most ultraviolet part of the fermion determinant is removed. We
define a reduced fermion matrix as \begin{equation}
\label{redfmat}
M(V)\: =\: M_{r}(V)A(V)\quad \mbox {with}
\end{equation}
\begin{equation}
\label{amat}
A(V)\: =\: \exp \left[ \alpha _{4}D^{4}(V)+\alpha _{2}D^{2}(V)\right] \, ,
\end{equation}
where \( D \) is the kinetic part of the fermion matrix and the parameters
\( \alpha _{4} \) and \( \alpha _{2} \) are arbitrary but real.
This way we achieve that an effective gauge action\begin{equation}
\label{effgaction}
S_{eff}(V)\: =\: -\frac{n_{f}}{2}\alpha _{4}Re\, tr[D^{4}(V)]-\frac{n_{f}}{2}\alpha _{2}Re\, tr[D^{2}(V)]
\end{equation}
is removed from the determinant. The action difference now has two
parts \begin{equation}
\label{DS}
\Delta S=S_{eff}(V)-S_{eff}(V^{\prime })+S_{Fr}(V)-S_{Fr}(V^{'}),
\end{equation}
where \( S_{Fr} \) is the fermionic action calculated with the reduced
fermionic matrix. For two flavors using an \( n \)th order polynomial
approximation the reduced fermionic action for the original smeared
fields is calculated as \begin{equation}
\label{SF_1}
S_{Fr}(V)=\overline{\Phi }^{\dagger }\overline{X}
\end{equation}
 with \[
\overline{\Phi }=A^{-1/2}(V)\, q_{-1/2}^{(n)}(M^{\dagger }(V)M(V))\, \xi \, ,\]
\[
\overline{X}=M^{\dagger }(V)M(V)\, \xi \, ,\]
or with an improved operator as \begin{equation}
\label{SF_2}
S_{Fr}(V)=\frac{1}{2}(3\overline{\Phi }^{\dagger }\overline{X}-\widetilde{X}^{\dagger }\widetilde{X})
\end{equation}
with \[
\widetilde{X}=P_{-1/2}^{(n)}(M^{\dagger }(V)M(V))\, \overline{X}\, .\]
 Here the vector \( \xi  \) is generated as\begin{equation}
\label{xi_nf2_r}
\xi =A(V^{\prime })^{1/2}\, q^{(m)\dagger }_{-1/2}(M^{\dagger }(V^{\prime })M(V^{\prime }))\, R,
\end{equation}
 where \( R \) is a Gaussian random vector. Note that in computing
the vector \( \xi  \) we use a polynomial of order \( m \) that
can be different from the order \( n \) used in computing the vectors
\( \overline{\Phi },\, \overline{X},\, \widetilde{X} \). For completeness
in eqs. (\ref{SF_1}-\ref{xi_nf2_r}) we again stated the argument
of the function \( q_{-1/2}^{(n)} \). In deriving these formulas
we used the fact that the matrices \( A \) and \( M_{r}^{\dagger }M_{r} \)
are functions of the matrix \( M^{\dagger }M \) and commute. The
vectors \( \xi ,\, \overline{\Phi },\, \overline{X},\, \widetilde{X} \)
are defined on the even sites of the lattice only. The reduced fermionic
action \( S_{Fr}(V^{\prime }) \) for the updated smeared fields is
calculated similarly and the proposed update is accepted or rejected
with the probability eq. (\ref{pacc}). The parameters \( \alpha _{2} \)
and \( \alpha _{4} \) in eq. (\ref{amat}) can be optimized to maximize
the acceptance rate. We keep the same choice \( \alpha _{2}=-0.18 \)
and \( \alpha _{4}=-0.006 \) as in \cite{Knechtli:2000ku}. It is
worth emphasizing that this algorithm works with smeared links only.
Smearing removes most of the ultraviolet fluctuations and the acceptance
probability of the heat bath and over relaxed algorithms is large
enough to make the algorithm efficient.

\section{The Performance of the Algorithm}

We have tested the above algorithm on \( 8^{3}\times 4 \) lattices
around the critical point of the \( N_{t}=4 \) phase transition with
quark masses \( am_{q}=0.01 \) and \( am_{q}=0.04 \). We found that
the two flavor simulation is considerably more efficient than the
four flavor simulation. One can update many more links (about twice
as many) in the over relaxation and heat bath steps without decreasing
the acceptance rate. The effectiveness of the algorithm did not change
as we lowered the quark mass. 

To test the accuracy of the polynomial approximation we calculated
the fermionic action eq. (\ref{SF_1}) and its improved version eq.
(\ref{SF_2}) using the same \( \xi  \) vector and compared the results
obtained with \( n=32,64,128 \) and 256 order polynomials. Using
the improved form eq. (\ref{SF_2}) we found very little difference
between the polynomials, even \( n=32 \) predicts the action difference
to an accuracy of 0.01 and the acceptance probability to an accuracy
of better than \( 10^{-3} \) on \( 8^{3}4 \) lattices. The simpler
form of eq. (\ref{SF_1}) gives similar results with \( n=64 \) or
higher order polynomials. On larger lattices higher order polynomials
might be necessary. On an \( 8^{3}\times 24 \), \( \beta =5.3 \),
\( am_{q}=0.04 \) lattice that has a lattice spacing \( a\approx 0.23 \)fm
and physical volume over to 30fm\( ^{4} \) we found it necessary
to use 64th order polynomials even with the improved form.

To check the accuracy of the \( \xi  \) vector we performed two long
runs using \( m=64 \) and \( m=128 \) order polynomials in eq. (\ref{xi_nf2_r})
at a coupling deep in the confining region on \( N_{t}=4 \) lattices
(\( \beta =5.0 \), \( am=0.01 \)). We found no difference between
the two runs within statistical accuracy.
\begin{figure}
{\centering \resizebox*{16cm}{!}{\includegraphics{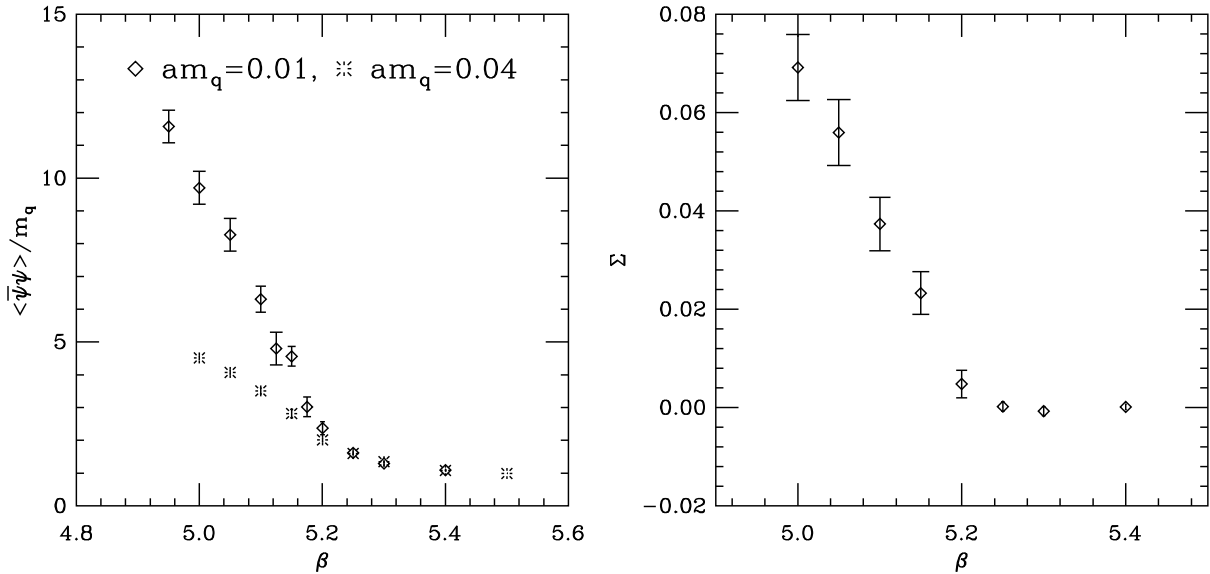}}  \par}

\caption{On the left, the chiral condensate \protect\( <\bar{\psi }\psi >\protect \)
divided by the bare quark mass is shown as a function of \protect\( \beta \protect \).
The results are from simulations on \protect\( 8^{3}\times 4\protect \)
lattices at two quark mass values. On the right, \protect\( \Sigma \protect \),
the condensate extrapolated to zero quark mass according to eq. (\ref{sigma_extr})
is plotted. \label{condensate}}
\end{figure}

\section{Finite Temperature Simulations}

We tested the algorithm presented in this article by studying the
finite temperature phase diagram of two-flavor QCD. Our first results
were obtained on \( 8^{3}\times 4 \) lattices for two values of the
bare quark mass, \( am_{q}=0.04 \) and \( am_{q}=0.01 \). We consider
the chiral condensate\begin{equation}
\label{psibarpsi}
\overline{\psi }\psi =\frac{1}{N^{3}_{s}N_{t}}\, trM^{-1}(V)\, ,
\end{equation}
 where \( N_{s} \) is the spatial lattice size. The chiral condensate
divided by the bare quark mass is shown in the left plot of Fig. \ref{condensate}.
The two mass values agree for \( \beta >5.2 \) but deviate at smaller
couplings. At small quark masses one expects that the chiral condensate
depends linearly on the quark mass \begin{equation}
\label{sigma_extr}
\frac{\left\langle \overline{\psi }\psi \right\rangle }{am_{q}}=\frac{\Sigma }{am_{q}}+c
\end{equation}
where \( \Sigma  \) is the value of the condensate extrapolated to
zero quark mass. Such an extrapolation is meaningful only if all extrapolated
points and \( am_{q}=0 \) are in the same phase. In that case \( \Sigma  \)
is zero in the chirally symmetric phase and finite, positive in the
chirally broken one. In the right plot of figure \ref{condensate}
we  show \( \Sigma  \) from our data. Since we have done simulations
with only two quark mass values we have no control over the quality
of the extrapolation nor can we be certain that our data points are
in the same phase as \( am_{q}=0 \). The plot nevertheless suggests
that for \( \beta >5.2 \) we are in the chirally symmetric phase
while for \( \beta <5.2 \) at least one of the mass values are in
the chirally broken phase. A more careful analysis with several quark
mass values must be done to resolve the phase diagram as has been
proposed in Ref. \cite{Hasenfratz:2001ef}. Nevertheless even with
this exploratory study it is evident that the two and four flavor
systems are very different. Further work to determine the phase diagram
and the corresponding scale of the two flavor system is under way.

\section{Acknowledgements}

We benefited from many discussions with Prof. T. DeGrand. We especially
appreciate that he drew our attention to Ref. \cite{Montvay:1996ea}
that describes the polynomial approximation. Prof. I. Montvay was
most helpful in answering our questions regarding the above reference
and for sharing his Maple code with us. F. Perez helped us write the
Mathematica code that we used to calculate the approximating polynomials
and their roots. This work was supported by the U.S. Department of
Energy. We thank the MILC collaboration for the use of their computer
code.

\bibliographystyle{apsrev.bst}
\bibliography{hyp}

\end{document}